# Singular skeleton evolution and topological reactions in edge-diffracted circular optical-vortex beams

Aleksandr Bekshaev[1*], Aleksey Chernykh[2], Anna Khoroshun[2], Lidiya Mikhaylovskaya[1]

[1]*Odessa I.I. Mechnikov National University, Dvorianska 2, 65082 Odessa, Ukraine*
[2]*East Ukrainian National University, Pr. Radiansky, 59-A, Severodonetsk, Ukraine*
[*]*Corresponding author: bekshaev@onu.edu.ua*

Edge diffraction of a circular optical vortex (OV) beam transforms its singular structure: a multicharged axial OV splits into a set of single-charged ones that form the 'singular skeleton' of the diffracted beam. The OV positions in the beam cross section depend on the propagation distance as well as on the edge position with respect to the incident beam axis, and the OV cores describe regular trajectories when one or both change. The trajectories are not always continuous and may be accompanied with topological reactions, including emergence of new singularities, their interaction and annihilation. Based on the Kirchhoff-Fresnel integral, we consider the singular skeleton behavior in diffracted Kummer beams and Laguerre-Gaussian beams with topological charges 2 and 3. We reveal the nature of the trajectories' discontinuities and other topological events in the singular skeleton evolution that appear to be highly sensitive to the incident beam properties and diffraction geometry. Conditions for the OV trajectory discontinuities and mechanisms of their realization are discussed. Conclusions based on the numerical calculations are supported by the asymptotic analytical model of the OV beam diffraction. The results can be useful in the OV metrology and for the OV beam's diagnostics.

## 1. Introduction

Diffraction is one of the most traditional and well-known phenomena of classical optics [1,2]. Of course, there are many quantitative details and special cases of diffraction that still need refinement and further elucidation but one may hardly expect that its thorough study can bring any peculiar news on the physical principles and general features of optical fields. However, this is not the case with structured light fields that have become a hot topic of modern optics during the past decades [3], especially, with light beams carrying optical vortices (OV) [4–6]. The edge diffraction of circular OV beams [7–20] shows many impressive non-trivial details associated with their special physical attributes: helical wavefront shape and transverse energy circulation. Even upon conditions of small diffraction perturbation (when the diffraction obstacle obscures just a far periphery of the beam cross section), the common and well studied diffraction effects (fringes, transverse diffusion of the light energy, etc. [1,2]) are supplemented with the OV-specific diffraction transformations.

Besides the asymmetric penetration of the light energy into the shadow region [9,13–15] impressively testifying for the transverse energy circulation in the incident beam, much attention was paid to the distribution and migration of the OV cores within the diffracted beam [7,8,11,12,14–18]. This interest is supported by the peculiar character of the OV cores as amplitude zeros and phase singularities, due to which they are physically highlighted and can be precisely detected and localized [21–23], which is employed, e.g., in the sensitive metrology [24–27]. In particular, a statistical technique for fringe analysis has been demonstrated in the detection of an optical vortex [23].

It is well established, both theoretically and in experiment, that after diffraction of an incident circular OV beam, the singularity shifts from its initial axial position, and an $m$-charged OV is decomposed into a set of $|m|$ secondary single-charged ones thus forming the 'singular skeleton' [6] of the diffracted beam. Upon the diffracted beam propagation, the OV cores move along intricate spiral-like trajectories [16,20] carrying distinct 'fingerprints' of the incident beam and its disposition with respect to the diffraction screen. The similar evolution of the singular skeleton can be observed in a fixed cross section of the diffracted beam when the screen edge performs a monotonous translation in the transverse direction towards or away from the beam axis [17–19].

However, the singular skeleton evolution is not limited by the 'smooth' migration of the secondary OVs within the diffracted beam 'body'. Generally, this process is accompanied by various topological reactions [4,6]: the OV disappearance and regeneration [7,8,10], emergence of new OVs, their annihilation, etc. Normally, such events occur at the beam periphery and are related with the diffraction fringes, etc. [15,16] but some sorts of topological reactions are intimately connected with the 'regular' OV migration and constitute its part [19]. Importantly, the progress of these reactions is highly sensitive to the incident beam properties and the diffraction conditions (e.g, the screen edge position or the propagation distance behind the screen plane), which had even caused their erroneous interpretation as the 'rapid OV migration' [18]. Therefore, in addition to the general physical interest, these topological events offer potentially valuable and prospective means for precise measurements and diagnostics of the OV beam's characteristics.

In this paper, we present an attempt of the systematic study of the topological discontinuities that occur in otherwise smooth trajectories of the OV migration in the optical fields obtained by means of the edge diffraction of circular Laguerre-Gaussian (LG) [4–6] and Kummer [28] vortex beams. We describe the typical manifestations of such discontinuities ('jumps') associated with the birth of the OV dipole at a remote point of the beam cross section followed by collision of one of the dipole constituents with the initial OV and their annihilation. The physical nature of this effect is explained with the help of a simple analytical model of the diffracted field formation based on interference of the incident beam and the edge wave [2] formed due to the incident field scattering by the screen edge. The analytical model is refined by means of the asymptotic analysis of the Fresnel–Kirchhoff diffraction integral mainly derived in our previous work [18] but additionally modified in this study. This enabled us to introduce the numerical criterion for the OV trajectory 'jumps' whose validity is demonstrated in several examples of the singular skeleton evolution in both basic situations: when the observation plane is fixed and the diffracted beam structure changes due to the screen edge translation ($a$-dependent evolution, see Sec. 3) and when the screen edge is fixed but the observation plane moves along the propagation direction ($z$-dependent evolution, see Sec. 4). The observed discontinuities are also interpreted based on the transverse projections of the smooth and continuous 3D vortex lines in the diffracted field. In the Appendices A and B, we present helpful illustrations of the jump mechanism and an additional type of topological reaction associated with the far-field pattern of the diffracted vortex beams.

**2. Description of the diffraction model**

We follow the general scheme of the vortex beam diffraction [17–19] (see Fig. 1). Let the incident monochromatic paraxial beam be described in the screen plane S by the slowly varying complex amplitude distribution $u_a(x_a, y_a)$; then in the observation plane at a distance $z$ behind S the diffracted beam complex amplitude can be found via the Kirchhoff-Fresnel integral

$$u(x,y,z) = \frac{k}{2\pi i z} \int_{-\infty}^{\infty} dy_a \int_{-\infty}^{a} dx_a \, u_a(x_a, y_a) \exp\left\{\frac{ik}{2z}\left[(x-x_a)^2 + (y-y_a)^2\right]\right\} \quad (1)$$

where $k$ is the radiation wavenumber; in any cross section, the electric field of the paraxial beam equals to $\text{Re}[u(x,y,z)\exp(ikz - i\omega t)]$ with $\omega = ck$, $c$ is the velocity of light.

We consider two types of the incident vortex beams. The first one is the Kummer beam that is usual in experimental practice [17,18] where an OV beam is formed from an initial Gaussian beam with the help of a special 'vortex-generating' element VG (see Fig. 1a) – a helical phase plate or a diffraction grating with groove bifurcation ("fork" hologram). In this case, the incident OV beam can be described [28], in the screen plane $(x_a, y_a)$, by the complex amplitude distribution $u_a(x_a, y_a) = u^K(x_a, y_a, z_h)$,

$$u^K(x_a, y_a, z_h) = \frac{z_{he}}{z_h}\sqrt{\frac{\pi}{2}}(-i)^{|m|+1}\exp\left[\frac{ik}{2z_h}(x_a^2 + y_a^2) + im\phi_a\right]\frac{z_R}{z_{he} - iz_R}e^{-A}\sqrt{A}\left[I_{\frac{|m|-1}{2}}(A) - I_{\frac{|m|+1}{2}}(A)\right]. \quad (2)$$

Here $z_h$ is the distance from the VG to the screen (see Fig. 1a), $\phi_a = \arctan(y_a/x_a)$ is the azimuth (polar angle) in the screen plane, $m$ is the OV topological charge (corresponds to the phase increment $2m\pi$ upon the round trip near the beam axis), $I_\nu$ denotes the modified Bessel function [29];

$$A = \frac{1}{4}\frac{z_R}{z_{he} - iz_R}\left[\frac{k}{z_{he}}(x_{ae}^2 + y_{ae}^2)\right], \quad z_R = kb^2, \quad (3)$$

$$z_{he} = \frac{z_h}{1 + z_h/R}; \quad x_{ae} = x_a\frac{z_{he}}{z_h}; \quad y_{ae} = y_a\frac{z_{he}}{z_h}, \quad (4)$$

$b$ being the Gaussian beam radius at the VG plane, see Fig. 1a. Eqs. (2) – (4) admit the non-planar wavefront of the initial Gaussian beam, $R$ is the wavefront curvature radius; equation for $z_R$ in (3) just formally coincides with the Raleigh range definition [2] because for finite $R$, $b$ is no longer associated with the beam waist.

Another beam type is the standard LG beam that is more suitable in theoretical analysis (for simplicity, we restrict our consideration by the modes with zero radial index). In this case $u_a(x_a, y_a) = u^{LG}(x_a, y_a, z_c)$ where [2,4,5]

$$u^{LG}(x_a, y_a, z_c) = \frac{(-i)^{|m|+1}}{\sqrt{|m|!}}\left(\frac{z_{Rc}}{z_c - iz_{Rc}}\right)^{|m|+1}\left(\frac{x_a + i\sigma y_a}{b_0}\right)^{|m|}\exp\left(\frac{ik}{2}\frac{x_a^2 + y_a^2}{z_c - iz_{Rc}}\right). \quad (5)$$

Here $\sigma = \text{sgn}(m) = \pm 1$, $b_0$ is the Gaussian envelope waist radius, $z_c$ is the distance from the waist cross section to the screen plane (see Fig. 1b), and $z_{Rc} = kb_0^2$ is the corresponding Rayleigh length [2]; the current beam radius $b_c$ and wavefront curvature radius $R_c$ in the screen plane are determined by known equations

$$b_c^2 = b_0^2\left(1 + \frac{z_c^2}{z_{Rc}^2}\right), \quad R_c = z_c + \frac{z_{Rc}^2}{z_c}. \quad (6)$$

Substituting (2) and (5) into (1) one can find the diffracted beam characteristics for arbitrary propagation distance $z$ and the screen edge position $a$. The OV core locations can then be easily identified as isolated intensity zeros, $|u(x,y,z)|^2 = 0$ [17], or as points in which different equiphase lines $\arg[u(x,y,z)] = \text{const}$ converge [15,16].

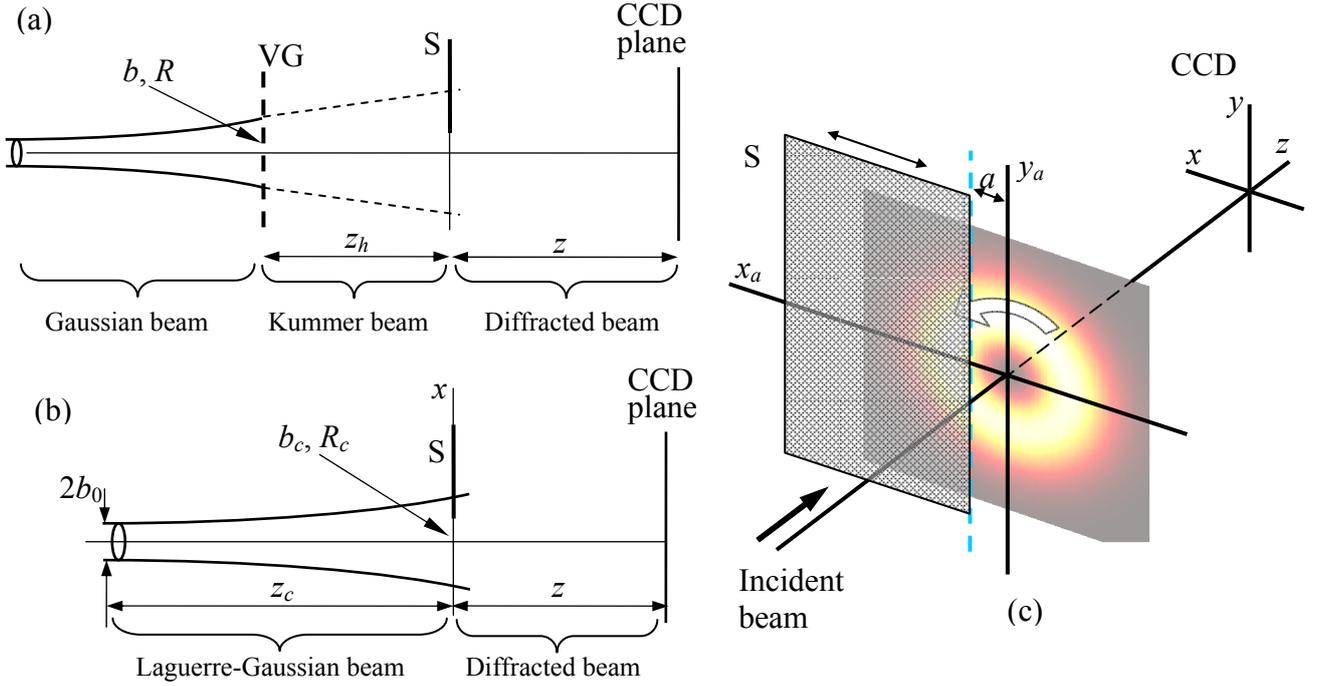

Fig. 1. Scheme of (a) formation and diffraction of the incident Kummer beam and (b) diffraction of the incident LG beam; (c) magnified view of the beam screening and the involved coordinate frames. VG is the OV-generating element, S is the diffraction obstacle (opaque screen with the edge parallel to axis *y*, its position along axis *x* is adjustable), the diffraction pattern is registered in the observation plane by means of the CCD camera. Further explanations see in text.

## 3. OV trajectories and their discontinuities

Examples of the OV trajectories within the diffracted beam cross section are presented in Fig. 2. These trajectories are calculated based on the numerical evaluation of the integral (1) for the experimental conditions of Ref. [17], i.e. for the Kummer beam (2) – (4) with $m = -3$ and

$$k = 10^5 \text{ cm}^{-1}, \quad b = 0.232 \text{ mm}; \quad R = 54 \text{ cm}, \quad z_h = 11 \text{ cm}. \qquad (7)$$

The images represent the patterns seen from the positive end of the *z*-axis (against the beam propagation). In panel (a), the lines of different colors indicate the constant-phase contours with increment 1 rad. Since the phase surfaces of singular beams are branched, they cannot be projected on the figure plane without cuts. These cuts are seen in the panel (a) as 'bundles' of lines of different colors merging together; each cut ends at an OV core. In Fig. 2a, three single-charged OVs are seen that originate from decomposition of the incident 3-charged OV due to the symmetry breakdown; Figs. 2b–d show the trajectories of OVs B – D, respectively.

Actually, Figs. 2b–d represent the refined and corrected results of Figs. 2c–e of Ref. [18]. We see the overall spiral-like motion complicated by radial pulsations, self-crossings, etc. Normally the spirals evolve oppositely to the energy circulation in the incident beam (cf. the grey curve in Fig. 2b) but locally a retrograde azimuthal motion takes place forming the 'loops'. Eventually, each OV migrates into the shadow region where it vanishes [7,8,17,18].

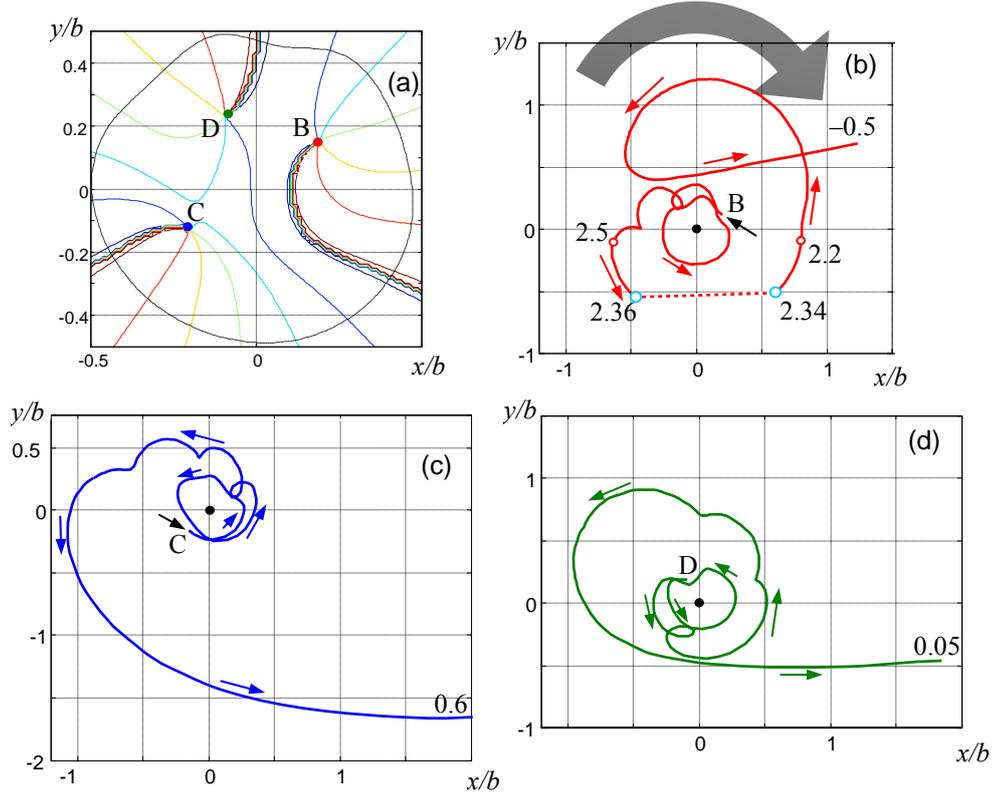

Fig. 2. Trajectories described by the OV cores in the cross section $z = 30$ cm behind the screen, the screen edge moving from $a = 4.4b$ to $a = -0.5b$ (see Fig. 1c), for the incident Kummer beam with topological charge $m = -3$ and parameters (7). The transverse coordinates are expressed in units of $b$ (7); large grey arrow shows the energy circulation in the incident beam (cf. Fig. 1c), small arrows show the directions of the OV motion. (a) 'Initial' positions of the three secondary OVs marked B, C and D for $a = 4.4b$, the thin black curve denotes the constant intensity contour at a level 10% of the maximum; (b) – (d) trajectories of OVs B, C and D while the screen edge advances (the final values of $a/b$ at which the corresponding OV disappears are marked near the ends of the curves), the beam axis is denoted by the black circle. The dotted line in panel (b) illustrates the OV "jump".

An important feature of the OV traces is that the OV motion along its trajectory is not uniform, which is most impressively evident in the trajectory of the OV B (Fig. 2b). While the screen performs a minute advance from $a = 2.36b$ to $a = 2.34b$, the OV abruptly 'jumps' between the points marked by cyan circles so that the trajectory looks apparently discontinuous (compare this with the adjacent trajectory segments where much larger screen shifts from $a = 2.5b$ to $a = 2.36b$ and from $a = 2.34b$ to $a = 2.2b$ cause noticeably smaller changes in the OV positions marked by the red circles). Also, while the OV B performs this 'jump', the positions of other OVs remain practically unchanged. In what follows, we intend to investigate the nature and mechanism of this effect.

### 3.1. Asymptotic analytical model

If the incident beam is an LG beam, the integral (1) can be, in principle, evaluated analytically but when $|m| > 1$, the analytic representation is cumbersome and physically obscure; for the incident Kummer beams the exact analytical representation is unknown. Nevertheless, the situation can be examined analytically by means of the simple model which is derived for $a \gg b$ but appears to be practically valid when the screen edge is separated by several *b* from the incident beam axis [18] (see Fig. 1). In this approximation, the diffracted beam (1) can be considered as a superposition of the unperturbed incident beam and the edge wave "emitted" by the screen edge [2]. For any circular OV beam considered in this paper, near the axis its complex amplitude distribution can be presented in the form

$$E_{inc} = B_0 \left(\frac{r}{b}\right)^{|m|} \exp(im\phi)\exp(ikz) \qquad (8)$$

where $r = \sqrt{x^2 + y^2}$ and $\phi = \arctan(y/x)$ are the polar coordinates in the observation plane. The quantity $B_0$ is a certain complex constant depending on the propagation distance and the beam type (e.g., Kummer or LG), as well as on its specific parameters, that can be easily derived from the explicit expressions (2) or (5). Near the origin of the observation plane, the edge-wave amplitude approximately amounts to

$$E_{edge} = D_0(a,z)\exp\left[ik\left(z + \frac{a^2}{2z} - x\frac{a}{z}\right)\right] = D_0(a,z)\exp\left[ik\left(z + \frac{a^2}{2z} - r\frac{a}{z}\cos\phi\right)\right] \qquad (9)$$

with the complex coefficient $D_0(a,z)$ that decreases with growing |*a*| and *z*. Eq. (9) differs from the similar expression used in Ref. [18] by the *x*-proportional term responsible for the wavefront inclination in the (*xz*) plane (see Fig. 1). Positions of the OV cores are determined by the condition $E_{edge} + E_{inc} = 0$, which entails

$$\frac{r}{b} = \left[\frac{|D_0(a,z)|}{|B_0|}\right]^{1/|m|}, \qquad (10)$$

$$\phi + M\cos\phi = C_{N+1} + k\frac{a^2}{2mz} \qquad (11)$$

where

$$M = \frac{kra}{mz}, \qquad (12)$$

and the coordinate-independent term $C_{N+1}$ possesses its own value for each secondary OV numbered by $N = 0, 1, \ldots |m|-1$,

$$C_{N+1} = \frac{1}{m}\left[\arg D_0(a,z) - \arg B_0 + (2N-1)\pi\right]. \qquad (13)$$

Despite their very approximate character, Eqs. (10) and (11) enable efficient qualitative analysis of the OV trajectories. First, one can note that under conditions of weak diffraction perturbation, the OV off-axis displacement $r \to 0$ and the second summand in the left-hand side of (11) can be neglected ($M \to 0$). Then Eq. (11), in full agreement with the experiment [17], predicts the monotonous behavior of the OV azimuth upon monotonous variation of *a* or *z*, which together with the monotonous nature of $D_0(a,z)$ in Eq. (10) dictates the spiral character of the OV trajectory. Also, Eq. (11) with $M \to 0$ makes it obvious that the rate of the OV spiral evolution should slow down with decrease of *a* and increase of *z*, which is also confirmed by experiments and numerical calculations [17,18].

However, the trajectory details we are studying in this paper appear at not very small *r* when the cosine term in (11) cannot be discarded. Then the azimuthal coordinate of the OV core is determined by the transcendent Eq. (11) which, in contrast to its counterpart of Ref. [18] cannot be solved analytically. Its qualitative analysis is illustrated by Fig. 3a. The left-hand side as a function of $\phi$ is imaged by the blue curve (for comparison, the thin light-blue line represents the left-hand side in the limiting case $M \to 0$), each horizontal line expresses a certain value of the right-hand side depending on $a$ and $z$ for a certain secondary OV number $N$. The solution $\phi(a,z)$ is obtained as an intersection of the blue curve and the corresponding horizontal line. In the 'normal' situation, $M \to 0$, there is only one intersection point (see, e.g., points $\phi_1$ and $\phi_4$ in Fig. 3a). When applied to the case of $m < 0$ presented in Fig. 2, with $a$ decreasing monotonically the horizontal line moves upward, and the corresponding $\phi(a,z) = \phi_1$ also changes monotonically and continuously. However, due to the trigonometric term in Eq. (11), the left-hand side can be non-monotonic, and at certain values of $a$ and $z$, the horizontal line reaches the region where the blue curve is nearly horizontal or decreases (e.g., between the red dashed lines in Fig. 3a, $\phi_2 < \phi < \phi_3$). Obviously, in this region $\phi(a,z)$ can change very rapidly; besides, there appear additional intersections that testify for nothing but emergence of additional OVs.

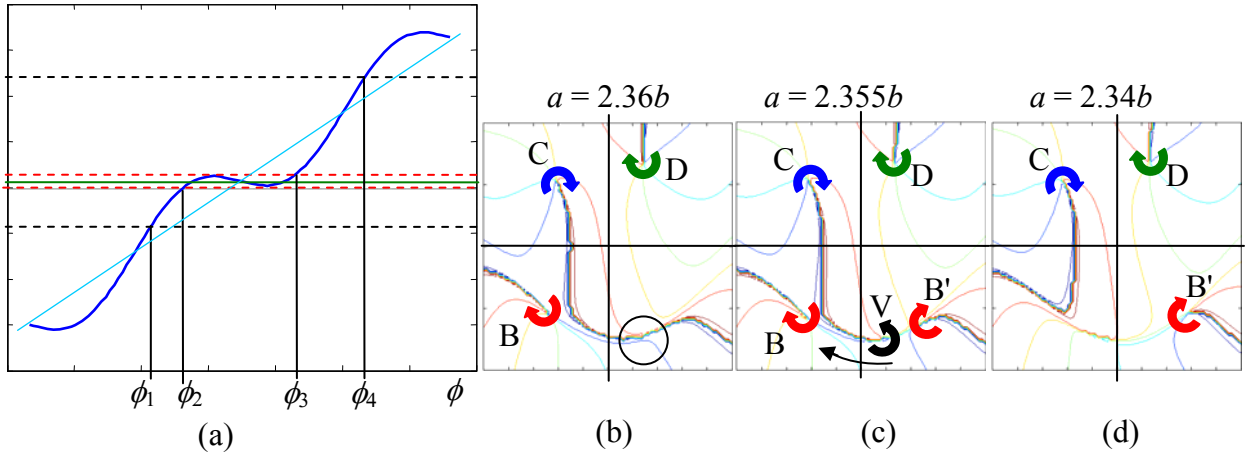

Fig. 3. (a) Illustration for the solution of Eqs. (11) and (15) (see also Video 1 [33]): The blue curve is the plot of the left-hand side expression for $|M| = 1.4$, horizontal lines symbolize different $(a, z)$-dependent values of the right-hand side. (b) – (d) Equiphase contours and the secondary OV positions in the cross section of the diffracted beam of Fig. 2 (see also Video 2 [33]); curve arrows show the local energy circulation near the OV cores; the screen-edge positions are indicated above each panel (further explanations in text).

3.2. The 'jump' description: Kummer beams

This procedure can be readily refined by employing the asymptotic representation of the diffracted beam field [18]; we only should take into account the linear $x$-dependent terms in the expression $P(x_a, x, d)$ (Eq. (A3) of Ref. [18]) that were discarded previously. So the second argument of $P(x_a, x, d)$ that was set 0 in Eqs. (A8), (A15) and (A18) of [18], should be restored and, accordingly, summands $-ik(ax/z)$ should be added to the exponents in brackets of Eqs. (7) and (19) of [18]. As a result, for the diffraction of the Kummer beam (2) – (4), instead of the simple relations (10), (11), the OV polar coordinates should be determined via equations (cf. Eqs. (14), (15) of [18])

$$\frac{r}{b} = \left\{ \left| \frac{D_1}{a^3} - \frac{D_2}{a} \exp\left[ \frac{ika^2}{2}\left( \frac{1}{z_d} - \frac{1}{z_h} \right) \right] \right| \cdot |B_1|^{-1} \right\}^{1/|m|}, \qquad (14)$$

$$\phi + M\cos\phi = \frac{1}{m}\arg\left[ \frac{D_1}{a^3}\exp\left( \frac{ika^2}{2z_h} \right) - \frac{D_2}{a}\exp\left( \frac{ika^2}{2z_d} \right) \right] + \frac{ka^2}{2mz} - \frac{1}{m}\arg B_1 + \frac{2N}{m}\pi \qquad (15)$$

where $B_1$, $D_1$ and $D_2$ are determined by Eqs. (8) – (10) of [18], $M$ is defined by (12) and

$$\frac{1}{z_d} = \frac{1}{z_{he} - iz_R}\left(\frac{z_{he}}{z_h}\right)^2 + \frac{1}{z_h + R} = \frac{i}{kb^2(z_h)} + \frac{1}{R(z_h)}, \quad (16)$$

$b(z_h)$ and $R(z_h)$ being the beam radius and the wavefront curvature radius which the initial Gaussian beam, incident onto the VG (see Fig. 1a), would have possessed in the screen plane if it had propagated "freely", without the VG-induced transformation (there was a mistake in the last equation of the Appendix of Ref. [18] that is now corrected in (16)).

The graphical solution of Eqs. (14) – (16) is illustrated by Video 1 [33] that shows evolution of the pattern of Fig. 3a for the Kummer incident beam with parameters (7), $m = -3$, $z = 30$ cm, while $a$ changes from $4.5b$ to $0.33b$; the three horizontal lines correspond to three secondary OVs with different $N$. In the Video 1, the evolution of the blue curve is more complicated than was discussed in the above paragraphs because of the variable $M$ (12), which depends on $a$ explicitly as well as implicitly, via $r$ and Eq. (14), and due to the more complex $a$-dependence of the right-hand side of Eq. (15); however, the principal details remain the same.

The existence of several intersections of the horizontal line with the blue curve (as for the green line in Fig. 3a) means that the smooth translational migration of the OV is no longer possible and is thus replaced by the topological reaction in which additional OVs emerge and annihilate [4]. Images of Figs. 3b–d and Video 2 [33] show the numerical example explaining the behavior of the OV B whose trajectory is depicted in Fig. 2b, within the 'jump' region. The OV positions are marked by the corresponding letters, as in Fig. 2b–d; additionally they are provided with curve arrows showing the local direction of the transverse energy circulation, colored in agreement with the trajectory colors in Fig. 2. While $a$ approaches the 'jump' region ($a = 2.36$ in Fig. 2b, point $\phi_2$ in Fig. 3a), there are three secondary OVs presented in Fig. 3b. At this moment, the small screen advance towards the axis almost does not affect the OV positions but induces a topological event: in the area indicated by the black circle in Fig. 3b, the cut is torn and the dipole of oppositely charged OVs emerges (see Fig. 3c). With further decrease of $a$, one of the new-born OVs, V, charged oppositely to all the other OVs (black curve arrow), rapidly moves against the 'normal' spiral OV motion. Then it meets the OV B and annihilates with it, whereas the second member of the dipole pair, B', still remains and starts its migration as a "continuation" of the OV B (Fig. 3d). Note that singularities C and D are practically stable during this process, and the 'virtual' OV V moves from B' to B along the smooth arc looking as a natural 'filling' of the spiral-like trajectory between $a = 2.36$ and $a = 2.34$. This agrees with the approximate Eq. (14) that dictates that radial coordinates of all OVs, including the 'virtual' ones, are determined by $a$ and $z$ independently of the azimuth $\phi$.

This example discloses the nature of the trajectory jump in Fig. 2b. It actually can be considered as a persistence of the same OV trajectory; however, within the 'jump' segment, a sort of the OV 'teleportation' occurs instead of the smooth translation.

The described anomalies of the OV trajectories in the diffracted beam are caused by the non-monotonic character of the left-hand side of Eq. (11) or (15), which takes place if the 'jump criterion' is realized,

$$|M| = \left|\frac{kra}{mz}\right| > 1, \quad (17)$$

and near the points where

$$\cos\phi = 0, \quad \frac{d}{d\phi}(M\cos\phi) < 0 \quad (18)$$

(the latter condition explains why the jump of Fig. 2b, as well as the noticeable acceleration of the OV motion in Figs. 2c, d [18] occur in the lower half-plane, near $\phi = 3\pi/2$; remember that $m < 0$ and, consequently, $M < 0$). In turn, Eq. (17) shows that the jump can preferably take place at large enough $a$ and not very high $z$; in particular, this explains why the numerical analysis reveals the 'jump' anomalies at $z = 30$ cm but they cannot be detected, with the same incident beam, at $z = 60$ cm and $z = 82$ cm [17,18]. In the present conditions of Eq. (7) and Fig. 2*b* with $z = 30$ cm, $a = 2.35b$, $r \approx 0.72b$, one finds $|M| \approx 1.01$, which agrees with the 'jump' existence. Noteworthy, the trajectories of the OVs C and D differ from the considered OV-B trajectory by the values of *a* and *r* at which they traverse the vicinity of $\phi = 3\pi/2$. For the OV C this occurs at $a = 3.75b$, $r \approx 0.25b$ (Fig. 2c), which gives $|M| = 0.56$; for the OV D – at $a = 3.1b$, $r \approx 0.4b$ (Fig. 2d) whence $|M| = 0.74$. This completely agrees with the absence of jumps and accompanying topological events in trajectories C and D.

### 3.3. Laguerre-Gaussian beams

According to the model of Sec. 3.1, the effects of 'jumps' in the OV trajectories within the diffracted beam cross section is common for any circular OV beam. We started its consideration with the special example of the Kummer beam where it was first noticed but the case of LG beam (5), (6) appears even more suitable for the general analysis. In this case, similarly to Eqs. (14) and (15), for large enough $a \gg b_c$, the OV coordinates can be described by approximate relations

$$r = \left[ a^{|m|-1} \exp\left( -\frac{a^2}{2b_c^2} \right) \left| \frac{D}{B} \right| \right]^{1/|m|}, \qquad (19)$$

$$\phi + M\cos\phi = \frac{1}{m}\left[ \arg D - \arg B \right] + \frac{ka^2}{2mz} + \frac{ka^2}{2mR_c} + \frac{2N}{m}\pi \qquad (20)$$

where *B* and *D* are determined by Eqs. (20) of Ref. [18], *M* is given by (12) (cf. Eqs. (21) and (22) of Ref. [18]). Note that it is the cosine term in the left-hand side of the azimuthal equation (20) that distinguishes Eqs. (19) and (20) from simplified Eqs. (4) and (5) of Ref. [20].

In Fig. 4, the numerically calculated OV trajectories for the diffracted multicharged LG beam (5) ($m = -3$) are presented. In the calculations we assumed the following values of the beam parameters:

$$k = 10^5 \text{ cm}^{-1}, \quad b_c = b_0 = b = 0.232 \text{ mm}, \quad z_c = 0, \quad R_c = \infty, \qquad (21)$$

that is, the beam waist coincides with the screen plane. As in the Kummer beam case (Fig. 2), there are three secondary OVs that evolve along the spiral-like trajectories and consecutively move to the shadow region where these vanish. The trajectories are marked by the same colors and the same letter notations as their counterparts in Fig. 2b–d. Generally, they show more regular and smooth behavior than in the case of Kummer beam, which is associated with the slower decay and oscillations of the Kummer beam intensity at $r \gg b$ [18,28]; remarkably, the analytical model of Eqs. (19), (20) give not only qualitative but also the fair quantitative characterization of the trajectory B even if $a \approx b$ (see Fig. 4a where the trajectory obtained analytically from Eqs. (19), (20) with $M = 0$ is presented as the thin dotted spiral; note that its final point corresponds to $a = 1.2b$).

Upon calculations, the 'jumps' were identified as events at which the additional pair of OVs emerge. For example, in Fig. 4a, while *a* decreases, the 'red' OV with topological charge −1 moves along the segment $B_0B$ and at the moment it approaches point B, the OV dipole is distinguished with −1-charged OV in point B'. This event takes place at $a = 1.98b$; then, the oppositely charged dipole member – 'virtual' OV V – rapidly moves along the black arc against the main spiral evolution. Meanwhile, the 'old' OV still continues its slow motion to meet the 'virtual' one until the annihilation occurs in the point marked by the circle at $a = 1.94b$. (Note that the 'virtual' OV distantly resembles the virtual particles in quantum theory [30]: it is short-living, and its only role is to implement the reaction transforming B into B'). During whole this process, the OV radial coordinate remains approximately constant, $r = 0.44b$. Similar events happen to the OV C at $a = 2.92b$ to $2.90b$ (Fig. 4b, $r = 0.27b$) and to the OV D at $a = 2.52b$ to $2.48b$ (Fig. 4c, $r = 0.35b$). In contrast to the situation of Fig. 2, now all the OVs experience rather articulate 'jumps', which is explained by the high values of the jump factor (17): $|M|$ = 1.56, 1.40 and 1.57 in cases of Fig. 3a–c, correspondingly.

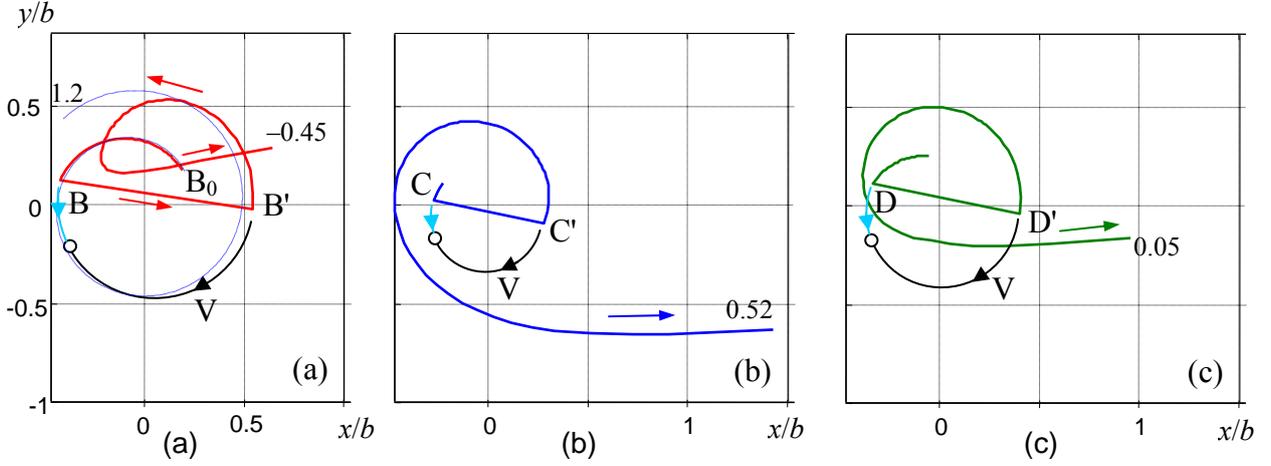

Fig. 4. Trajectories described by the OV cores in the cross section $z$ = 10 cm behind the screen, the screen edge moving from $a = 3b$ to $a = -0.45b$ (see Fig. 1c), for the incident LG beam with topological charge $m = -3$ and parameters (21). Each panel shows the trajectory of a single OV with additional explaining details. The transverse coordinates are expressed in units of $b$ (21), small arrows show the directions of the OV motion; the final values of $a/b$ at which the corresponding OV disappears are marked near the ends of the curves. The trajectories experience 'jumps' between points B and B', C and C', D and D', respectively; the black (cyan) arcs represent the motion of 'virtual' ('old') OVs before their annihilation in points marked by circles. In panel (a), the trajectory calculated analytically via Eqs. (19), (20) for $3b > a > 1.2b$ with $M = 0$ is depicted by the thin dotted curve for comparison.

## 4. OV jumps in the propagating diffracted beam

We have considered several examples in which the migration of the secondary OVs across a fixed cross section of the diffracted OV beam, caused by the screen edge advance, has been addressed. However, there is another interesting aspect of the singular skeleton evolution associated with its 3D nature: for a given screen edge position, the OV coordinates change with the observation plane distance $z$ [15,16,20]. According to the general physical arguments specified by the analytical suggestions supplied by Eqs. (10), (11), (14), (15) (19) and (20), the discussed mechanisms determining the OV trajectories are still in charge for the $z$-dependent evolution, and the trajectory discontinuities and topological reactions of the above-described type are expected to occur in this situation as they do in the $a$-dependent trajectories studied in Sec. 3.

### 4.1. Kummer beams

Fig. 5 represents the $z$-dependent evolution of the secondary OVs in the same diffracted beam that was analyzed in Sec. 3, 3.2 and 3.3 but for the fixed screen-edge position $a = 4b$ illustrated in the panel (a). Note that, to make the beam structure better visible, the transverse amplitude distribution $|u^K(x_a, y_a, z_h)|$ is presented instead of the more common intensity $|u^K(x_a, y_a, z_h)|^2$. Anyway, the screen barely 'touches' the beam periphery, which, nevertheless, induces quite observable and rich of details perturbations of its singular skeleton displayed in Fig. 5b–d. In case of a propagating beam, there always is present the trivial component of the OV migration associated with the overall beam divergence; to abstract from this non-informative component, in Fig. 5b–d the OV trajectories are displayed in the normalized transverse coordinates

$$x_e = x\left(1+\frac{z}{R}\right)^{-1}, \quad y_e = y\left(1+\frac{z}{R}\right)^{-1}. \tag{22}$$

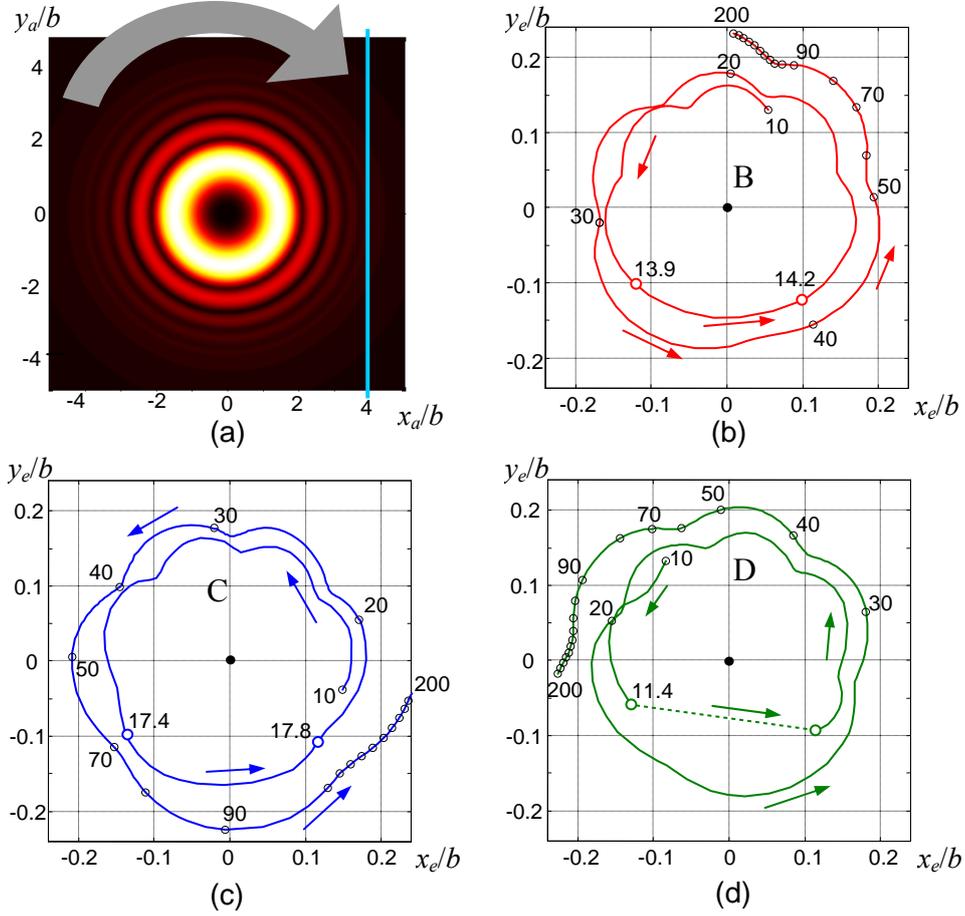

Fig. 5. Transverse projections of the OV trajectories behind the screen whose edge is fixed at $a = 4b$ (see Fig. 1c), for the incident Kummer beam with topological charge $m = -3$ and parameters (7) (cf. Fig. 2). (a) The screen edge position (blue line) against the incident beam amplitude distribution in the screen plane, the large arrow shows the energy circulation direction. (b) – (d) Separate OV trajectories for $z$ growing from 10 cm to 200 cm, letters B, C and D denote the same secondary OVs that are shown in Fig. 2; thin black empty circles correspond to $z$ values multiple of ten in centimeters, some of them are provided with corresponding numerical marks; colored white-filled circles mark the segments of rapid evolution. The horizontal and vertical coordinates are in normalized units of (22); small arrows show the directions of the OV motion. The trajectory 'jump' is seen only in panel (d) at $z = 11.4$ cm (dotted line).

In general, the OV trajectories of Fig. 5b–d are similar to those of Fig. 2b–d and show the same character of pulsating spirals. In the course of the beam propagation (growing $z$), the pulsation period increases and in the far field the pulsations vanish. In contrast to the trajectories of Fig. 2, here are no self-crossings ('loops' as in Figs. 2b–d); the apparent self-crossings near $z = 20$ cm in Fig. 5d are seeming and appear only in the normalized coordinates (22). The most important is that in case of the $z$-dependent evolution there also exist regions of very rapid OV migration (the trajectories' segments between the white-filled circles). In full agreement with the model of Sec. 3.2 (see Eq. (18) and Fig. 3a), these regions are in the lower half-plane (near the OV core azimuth $\phi = 3\pi/2$). However, the 'true' jump only happens to the OV D in the panel (d). This agrees with the criterion (17) that can be checked based on the presented trajectories: in Fig. 5b, $r = 0.18b$, $z = 14$ cm, and $|M| = 0.97$; in Fig. 5c, $r = 0.226b$, $z = 17.4$ cm, and $|M| = 0.93$; and only in Fig. 5d $r = 0.171b$, $z = 11.4$ cm, $|M| = 1.08$ – the conditions for the jump are realized, and it is indeed observed.

### 4.2. Laguerre-Gaussian beams

Diffraction of an LG beam provides additional and rather conspicuous illustrations for the 3D singular skeleton evolution [20]. Like in Sec. 3.3, we consider the incident LG beam (5) with its waist in the screen plane and the Gaussian envelope parameters (21) but with the topological charge $m = -2$ (Fig. 6). Despite that the chosen screen edge position $a = 2b$ can hardly be treated as a far periphery of the incident beam profile and the expected perturbation of its structure is rather strong, the OV migration looks remarkably regular (Fig. 6b).

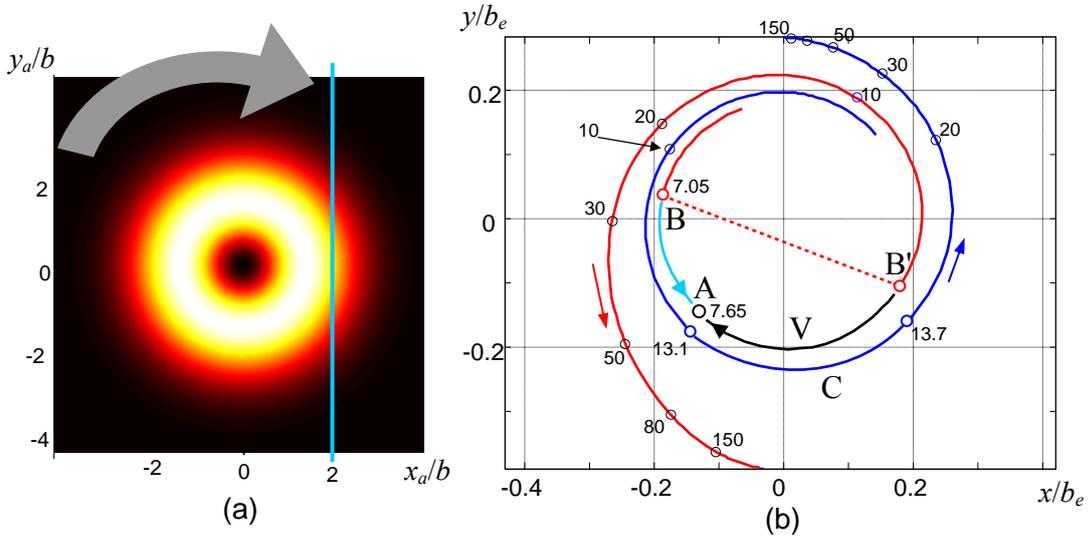

Fig. 6. Transverse projections of the OV trajectories behind the screen whose edge is fixed at $a = 2b$ (see Fig. 1c), for the incident LG beam with topological charge $m = -2$ and parameters (21). (a) The screen edge position (blue line) against the incident beam amplitude distribution in the screen plane, the large arrow shows the energy circulation direction. (b) Red (B) and blue (C) curves represent the trajectories of the two secondary OVs for $z$ growing from 5.6 cm to 530 cm ($9.85z_{Rc}$); black empty circles denote the intermediate $z$ values (marked in centimeters); colored white-filled circles mark the segments of rapid evolution The transverse coordinates are given in units normalized by (23); small arrows show the directions of the OV motion. At $z = 7.05$ cm, the OV B experiences the 'jump' into B' position shown by the dotted line; the cyan and black arcs represent the evolution of the 'old' B and of the 'virtual' OV V after the jump until they annihilate in the point A marked by the black empty circle (cf. Fig. A1 and Video 3, 4 [33]).

As in Fig. 5, to remove the trivial migration component associated with the beam divergence, the coordinates are normalized by the Gaussian envelope radius of the supposed unperturbed incident beam,

$$b_e = b\sqrt{1 + \frac{z^2}{z_{Rc}^2}} \qquad (23)$$

where, in view of Eq. (21), $z_{Rc} = z_R = 53.8$ cm is the Rayleigh length of the incident beam. Again, as in comparison of Figs. 4 and 2, the OV trajectories in the diffracted LG beam form almost perfect spirals, without pulsating irregularities observed in Figs. 5b–d for the diffracted Kummer beam. This difference between the singular skeleton patterns in Figs. 5b–d and Fig. 6b is most probably caused by the ripple structure [28] well seen in Fig. 5a: with growing off-axial distance $r_a$ the Kummer beam profile evolves in the oscillatory manner and its amplitude decreases very slowly at the beam periphery ($\sim r_a^{-2}$ instead of the exponential decay in an LG beam). The edge wave (9) is formed as a superposition of partial waves scattered by each point of the screen edge, and these waves obtain oscillating amplitudes and initial phases, in agreement to the oscillating behavior of the incident wave amplitude and phase along the screen edge. Accordingly, the edge wave complex amplitude $D_0(a,z)$ acquires the non-monotonous dependence on $a$ and $z$ which entails the non-monotonous behavior of the OV radial displacement (10). In case of the smooth transverse decay of the incident beam amplitude, the pulsations in the diffracted-beam OV trajectories vanish, as is seen for LG beams in Figs. 4 and 6b; the similar smoothening is expected for the incident Kummer beams (2) – (4) with large enough $z_h$ [28].

In Fig. 6b the OV B trajectory (red) experiences the jump at $z = 7.05$ cm while the OV C (blue) only shows the rapid evolution between $z = 13.1$ cm and $z = 13.7$ cm. This, again, is in full compliance with the criteria (17) and (18): for the OV C, $r = 0.234b$, and with $m = -2$, $a = 2b$, $z = 13.1$ cm this entails $|M| = 0.96$ whereas for the OV B, $r = 0.191b$, $z = 7.05$ cm, and $|M| = 1.46$. The jump mechanism is completely the same as in other examples: the OV dipole is born in point B' after which its oppositely charged 'virtual' member V rapidly moves 'backward' towards the 'old' B and annihilates with it in point A corresponding to $z = 7.65$ cm. This example supplies a spectacular dynamical illustration of the topological reactions and the 'virtual' OV migration accompanying the jump, which are presented in Appendix A, Fig. A1 and Videos 3, 4 [33].

### 4.3. 3D trajectories and the nature of discontinuities

To elucidate in more detail the discontinuous trajectory of the OV B in Fig. 6b, we present it as a 3D graph together with the trajectories of the 'old' OV B after the jump and of the virtual OV (cyan and black curves of Fig. 6b). The result given in Fig. 7 reveals that the three trajectories of Fig. 6b are actually fragments of the single 'full' curve that is perfectly continuous and smooth, so the jumps and topological reactions appear only in its projections (in particular, the red, cyan and black curves of Fig. 6b are projections of the corresponding segments of the curve of Fig. 7 viewed from the positive end of axis $z$). This agrees with the usual concepts of the OV filaments [6,31,32] and discloses the nature of the intriguing effects considered in previous sections.

Let the 'full' OV trajectory of Fig. 7 be represented in the parametrical form, i.e. the coordinates of a current trajectory point are expressed as functions of the trajectory length $s$ measured from the starting point at $z = 5.6$ cm:

$$x_v = x_v(s), \quad y_v = y_v(s), \quad z_v = z_v(s). \tag{24}$$

In a given transverse plane, the OV position is determined as an intersection between the plane and trajectory. The 'normal' evolution implies that everywhere $dz_v/ds > 0$, and then in each observation plane, only one intersection point can exist, but in some configurations of the diffracted beam singular skeleton, regions of a 'retrograde' evolution, where

$$dz_v/ds < 0, \tag{25}$$

may occur. It is such a situation that is depicted in Fig. 7 between the planes $P_1$ and $P_2$. When the observation plane approaches $P_1$ from the left, it 'touches' the trajectory at the additional point B' (a local minimum of the function $z_v(s)$), which corresponds to the dipole emergence. With further advance, the observation plane will contain three intersection points with the curve, which are interpreted as the 'teleported' OV B', 'old' OV B and the 'virtual' oppositely charged OV V. In the position $P_2$ the observation plane again touches the trajectory, now in point A with the local maximum of $z_v(s)$, and the intersections corresponding to B and V disappear: the two OVs annihilate.

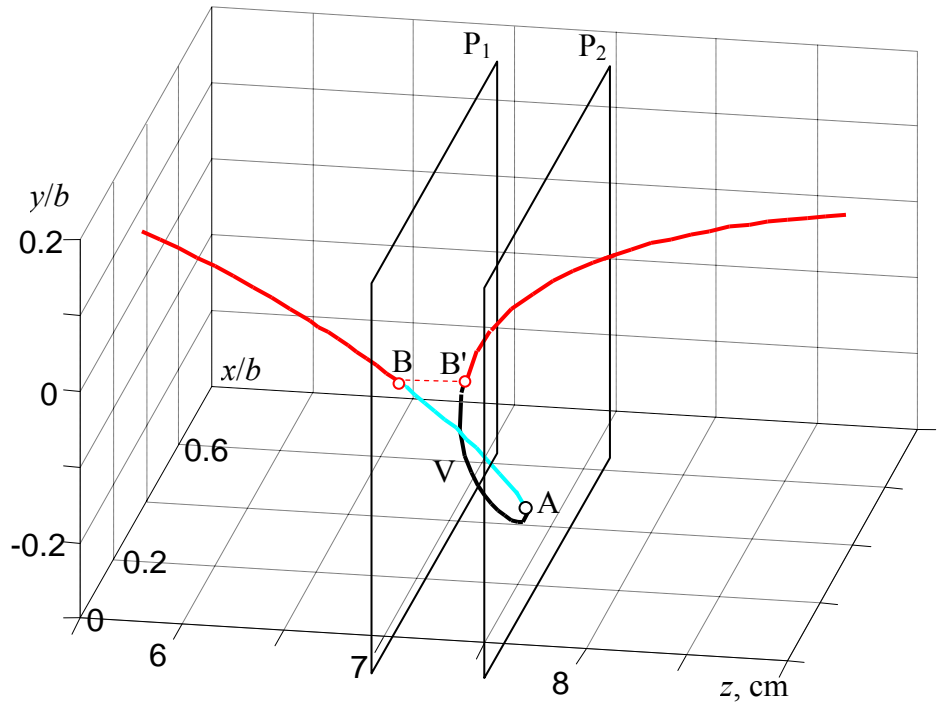

Fig. 7. 3D trajectory of the 'red' (B) OV of Fig. 6b (incident LG beam with $m = -2$ and parameters (21), screen edge position $a = 2b$) in the near-jump region (5.6 cm $< z <$ 9 cm). The transverse coordinates are given in units of $b$ (21); plane $P_1$ ($z = 7.05$ cm) crosses the trajectory in point B and is tangent to it in point B', plane $P_2$ ($z = 7.65$ cm) is tangent to the trajectory in the annihilation point A (black empty circle); the red, cyan and black segments correspond to the red, cyan and black arcs in Fig. 6b.

This picture completely explains the discontinuous trajectories of the OV cores not only in case of the $z$-dependent evolution (Sec. 4.1, 4.2) but also in case of the screen edge translation (Sec. 3, 3.2, 3.3). In the latter situation, the observation plane is fixed but the 'full' 3D curve is smoothly deformed with variation of $a$, and the 2D trajectory jump takes place if in the observation plane the condition (25) becomes true. In fact, the 'jump criterion' (17) is equivalent to (25), and this is why it is equally applicable to both the $z$-dependent and $a$-dependent variations of the diffracted beam singular skeleton.

Here we are nearly touching the aspect in which the theory of OV diffraction becomes entangled into the rich and stimulating field of the vortex lines and their geometry (see, e.g., [6] and references therein). This aspect deserves a special investigation; now we only remark that the intricate and at first glance artificial patterns of the OV lines that are deliberately generated by means of special procedures [6,32] can naturally exist in the edge-diffracted circular OV beams.

## 5. Conclusion

To summarize the main outcome of the paper, we underline that the observed and predicted peculiar details of the singular skeleton behavior are rather common for light beams with well developed singular structure, e.g. speckle fields [4,6]. In this view, the diffracted OV beams can be considered as their simplified models and, possibly, efficient means to create controllable singular-optics structures with prescribed properties, which can be useful in diverse research and technology applications.

In particular, the presence of the well developed, regular and easily interpretable singular structure makes the diffracted OV beams suitable objects for the general study of the OV lines and their geometric regulations, evolution of individual singularities, their transformations, topological reactions and interactions. On the other hand, the OV trajectories' discontinuities, 'jumps', birth and annihilation events described in this paper are, as a rule, highly sensitive to the incident beam parameters and the diffraction conditions. For example, the OV positions in the diffracted beam cross section can be sensitive indicators of the screen edge position with respect to the incident beam axis, which can be employed for precise distant measurements of small displacements and deformations [15,16]. From Figs. 2b and 4 one can easily see that near the 'threshold' conditions of topological reactions the screen edge displacement of $0.01b$ induces a two orders of magnitude larger OV jump in the diffracted beam. Note that such sensitivity is predicted without any special consideration; undoubtedly, a detailed analysis aimed at the search of the diffraction parameters most favorable for the distant metrology will improve these figures. This aspect of the present work enables to suggest its applications for the problems of the precise OV metrology [24–27] as well as for the incident OV diagnostics, which can be prospective in the fields of laser beam shaping and analysis and in optical probing systems.

It should be noted that the topological peculiarities discussed in this paper take place, as a rule, under conditions of a rather weak diffraction perturbation (the screen edge distance from the beam axis $a$, in any case, exceeds the incident beam radius $b$), and at rather small propagation distances $z$ (this follows from the 'jump' criterion (17)). In such situations, the diffraction-induced variations of the singular skeleton (e.g., displacements of the OV cores from the nominal beam axis) would presumably be small, and corresponding questions about their detectability may arise. However, according to Figs. 2 – 6, in the most interesting ranges of $a$ and $z$ these displacements reach several tenths of the incident beam radius, which is quite available for the precise measurement techniques.

Most of the quantitative results of the paper are obtained numerically but their interpretation is based on the asymptotic analytical model of Eqs. (11) – (13) with refinements (14), (15) and (19), (20). Remarkably, the model derived for the condition $a \gg b$ appears to be valid in the much larger and physically interesting domain; at least, for the LG beam diffraction it does not fail even at $a \approx 2b$, and the model-based criterion (17) works perfectly well in all the considered examples. However, the model predicts monotonic behavior of the OV radial displacement $r$ with growing $z$ for Kummer beams, i.e. does not explain the radial pulsations of the spirals in Fig. 5b–d. Nevertheless, we hope that despite its approximate character, the model will give a reliable analytical basis for further research of the vortex beams' diffraction. At least, all the conclusions concerning the spiral-like character of the OV trajectories and their jumps when the criteria (17) and (18) are satisfied, are absolutely reliable and supported by experiment [17]. The fine details of the OV trajectories in diffracted Kummer beams (self-crossings and pulsations in Figs. 2b–d and 5b–d), that appear due to the slow fall-off of the Kummer beam amplitude, are expected to be sensitive to the incident beam behaviour at the far transverse periphery. In this view, even the 'routine' approximations usually employed in the numerical simulations can be sources of errors, e.g., the integration domain limitation in the Fresnel-Kirchhoff integral (1). In such situations, the explicit allowance for the specific conditions of the Kummer beam preparation and for the optical system it passes would be necessary.

A possible direction of further research can be related with the more full characterization of the separate OVs in the diffracted beam. So far we were only interested in their positions; but no less informative can be their morphology and anisotropy parameters [5,6]: the orientation and the axes ratio of the equal-intensity ellipses in the nearest vicinity of the OV core. Especially, under conditions close to topological reactions, the OVs are highly anisotropic, and this supplies additional markers to characterize the qualitative discontinuities in the singular skeleton evolution.

Another way of possible further development of ideas and approaches introduced in the present paper can be oriented at the search of special conditions of the OV beam preparation and diffraction, which provide especially high sensitivity for the metrological and diagnostic applications outlined two paragraphs above.

Finally, we note that the approach developed in this paper can be extended to more complicated cases of the OV diffraction, e.g. when the diffraction 'obstacle' can be modelled by an inhonogeneous transparency with complex transmission function depending on one of the transverse coordinates, $T(x) = t(x)\exp[i\varphi(x)]$ with real $t(x)$ and $\varphi(x)$, $|t(x)| \leq 1$. Then the diffraction problem can be reduced to the analogue of Eq. (1) where the upper limit of the inner integral is infinite but for $x > a$ the integrand function is $T(x)u_a(x_a, y_a)$. For the case of small diffraction perturbation, $a \gg b$, approximate expressions similar to Eqs. (14), (15) and (19), (20) can be derived, which implies that the main qualitative features of the OV migration in the output beam cross section (spiral-like trajectories and 'jumps') will again take place. However, the quantitative parameters of the trajectories (off-axial OV displacements, magnitude and articulateness of 'jumps', etc.) will depend on the magnitude and abruptness of the transparency-induced transformations. This interesting and important problem goes beyond the scope of the present work but will be a task of the special future investigation.

**Appendix A. The 'jump' dynamics**

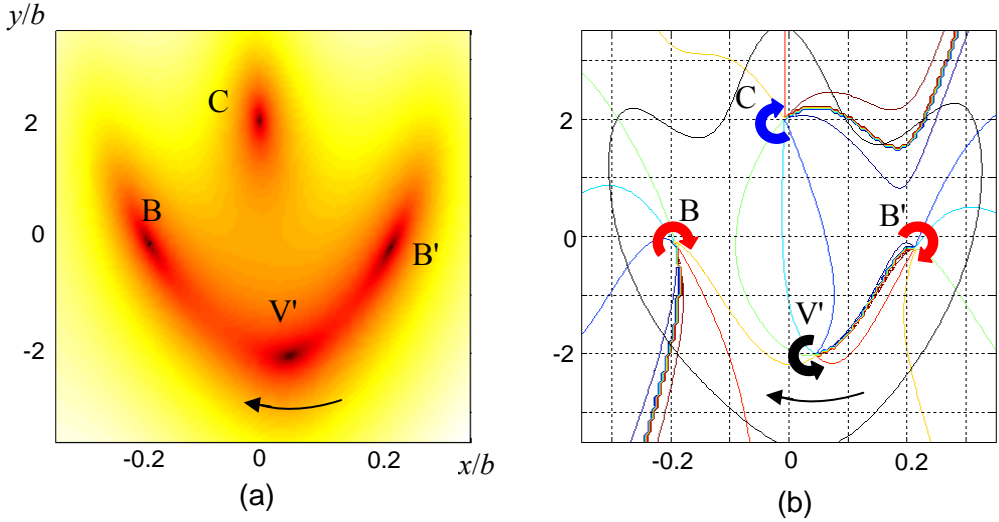

Fig. A1 (see also Video 3 and 4 [33]). Near-axis intensity and phase distributions in the diffracted LG beam of Fig. 6 ($m = -2$, $a = 2b$) in the cross section $z = 7.35$ cm (between planes $P_1$ and $P_2$ in Fig. 7); the transverse coordinates are in units of $b$ (21). (a) Pseudocolor map of the transformed intensity distribution (A1) with enhanced visibility of the amplitude zeros; dark spots are the OV cores; (b) Equiphase contours (colored), the thin black curve denotes the constant intensity contour at the level 10% of the maximum; curve arrows show the local energy circulation in the vicinity of the OV cores marked conventionally as in Fig. 6b. B is the 'old' OV (remainder of the 'red' OV evolution for $z > 7.05$ cm, cf. the cyan arc in Fig. 6b), B' is its continuation after the jump (negatively charged member of the newborn dipole). The oppositely charged 'virtual' OV V (black curve arrow) moves from B' to B (thin arrow), the OV C (blue) remains stable.

This presentation shows the evolution of the diffracted beam transverse profile for the incident LG beam considered and discussed in Sec. 4.2, 4.3, Figs. 6 and 7 (topological charge $m = -2$, plane wavefront, screen-edge position $a = 2b$) within the 'jump' region 7.0 cm $< z <$ 7.8 cm. Fig. A1 demonstrates the momentary 'snapshot' of this evolution at $z = 7.35$ cm. To enlarge the contrast in the low-intensity area, Fig. A1a and Video 3 [33] represent the transformed intensity distribution [17]

$$I_T(x, y) = \left[|u(x, y)|^2\right]^{1/15}. \tag{A1}$$

Both the intensity (Fig. A1a, Video 3 [33]) and phase (Fig. A1b, Video 4 [33]) clearly demonstrate the mechanism of the OV jump which is, in essence, the same as in case of a fixed diffracted beam section and varying screen position $a$ (Fig. 3c–d, Video 2 [33]) but the corresponding processes and topological reactions look even more impressive. At a certain distance of propagation $z$ ($z = 7.05$ cm in our example), in a certain point remote from the OV B, the OV dipole (B', V) emerges. The dipole member B' with the same sign as the incident OV moves slowly in agreement to the general spiral evolution while the oppositely charged dipole member V (black curve arrow in Figs. 6b and A1b) rapidly moves against the spiral evolution to meet the 'old' OV B and eventually annihilates with it. The OV B' continues the 'regular' spiral motion.

**Appendix B. Topological reactions in the diffracted beam far field**

Based on several exampled of the diffracted OV beams' behavior, it was established in Ref. [16] that when the incident LG beam has the plane wavefront, in the far field ($z \to \infty$) all the OVs are concentrated on the axis parallel to the screen edge, i.e. on the vertical axis in our case. This rule is fulfilled in Fig. 6 but is apparently violated in Fig. 5. This can be attributed to the fact that Fig. 5 illustrates the evolution of the diffracted Kummer beam rather than the LG one, and that its wavefront at the screen plane is not plane, but, anyway, it is remarkable that Figs. 5b–d show no tendency of the OVs arranging along any straight line with growing $z$. This observation is confirmed by an example of the diffraction of the incident LG beam with $m = -3$ and parameters (21) (see Fig. B1a).

In fact, this is the same beam that is considered in Sec. 3.3 and Fig. 4 but now the screen edge position is fixed, $a = 3b$, and the singular skeleton evolution with increasing $z$ is illustrated. In Fig. B1a, in contrast to Figs. 5b–d and 6b and to make the difference in the separate OVs' azimuthal positions more impressive, the transverse OV coordinates are deliberately not normalized by any $z$-dependent multiplier like (22) or (23), and the trajectories demonstrate the real 'radiant' OV migration. Their far-field azimuthal coordinates obviously tend to

$$\phi_C = \frac{3\pi}{2}, \quad \phi_D = \frac{3\pi}{2} - \frac{2\pi}{3}, \quad \phi_B = \frac{3\pi}{2} - \frac{4\pi}{3}. \tag{B1}$$

Note that the analytical model (20) for $z \to \infty$ and with account for (21) just predicts $\phi_N = -\frac{\pi}{2} + \frac{2N\pi}{m}$, which agrees with Eq. (B1) and Fig. B1a.

Here is an evident contradiction to the conclusions of Ref. [16], which can only be explained by that the previous consideration [16] was restricted to the situations of a rather severe screening, $a < 1.0b$. That is, a certain transition from the 'radiant' far-field OVs' distribution of Fig. B1a to their arrangement along the vertical axis, like in Fig. 6b, should take place when $a$ changes from $3b$ to $b$. And this is really so. With decreasing $a$, Eqs. (20) and (B1) are no longer valid but the numerical study shows that the OV C of Fig. B1a continues its off-center motion along the lower vertical half-axis whereas the OVs B and D approach symmetrically the upper half-axis until they meet each other.

The final stages of this process, when the screen advances from $a = 1.4b$ to $a = 1.16b$, are illustrated by Video 5 [33] and Fig. B1b–d; for convenience, the far-field coordinates $\theta_{x,y} = (x, y)/z$ are expressed in units of the incident Gaussian envelope self-divergence angle

$$\gamma = (kb_0)^{-1}. \tag{B2}$$

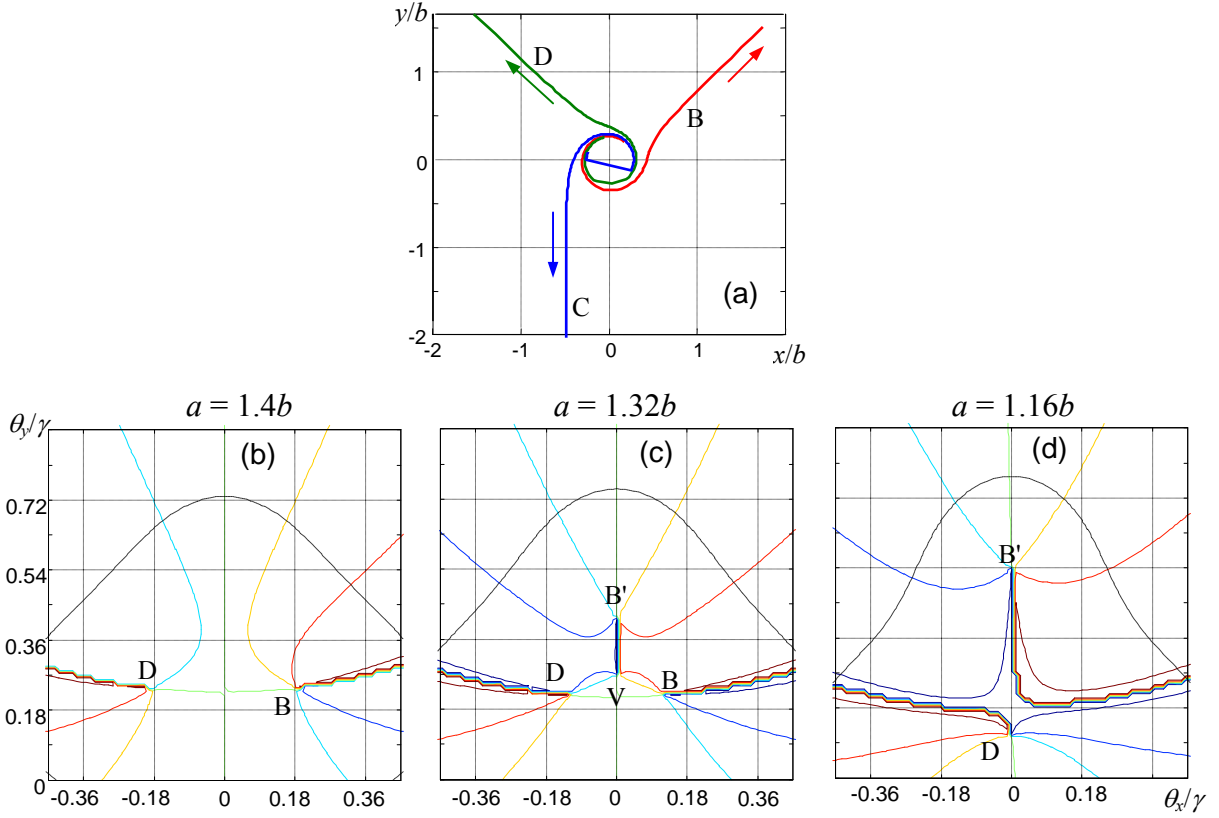

Fig. B1. (a) Transverse projections of the trajectories described by the OV cores upon the diffracted beam propagation from $z = 10$ cm to 400 cm, for the incident LG beam (5) with $m = -3$ and parameters (21) and the fixed screen edge position $a = 3b$ (cf. Sec. 3.3 and Fig. 4). The transverse coordinates are in units of $b$ (21), small arrows show the directions of the OV motion (B, C and D in correspondence to Fig. 4). (b) – (d) Equiphase contours and the OV positions in the far-field cross section $z \to \infty$ for varying screen edge position, the transverse angular coordinates are expressed in units of $\gamma$ (B2): (b) $a = 1.4b$, (c) $a = 1.32b$ and (d) $a = 1.16b$. OVs B, D and C correspond to the identically marked OVs in Fig. B1a and Fig. 4; with the screen edge advancing to the axis, the OV dipole (B', V) is formed, V moves downwards and annihilate, and finally B' and D remain on the vertical axis (see Video 5 for details).

It is seen that here, again, the topological reactions take place. While the OVs B and D get close to the vertical axis (at $a = 1.34b$), an OV dipole (B', V) emerges exactly on the vertical axis (Fig. B1c shows the situation when the dipole is already well developed and its members are distinctly separated). With the further screen advance, one of the new-born OVs, B', moves off-center along the vertical axis whereas the second one – the 'virtual' oppositely charged vortex V – approaches the pair B, D. Finally, at $a = 1.26b$ the topological reaction between the two $-1$-charged OVs B, D and the $+1$-charged OV V takes place, which results in the single negative OV that remains on the vertical axis and slowly moves downward with further decrease of $a$ (in Fig. B1d it is marked D conventionally but in fact, the ternary topological reaction takes place in which the 'input' OVs B, D and V equally contribute to produce the new 'output' one that remains attached to the vertical axis).

This reconciles our new results of Figs. 5b–d and Fig. B1a with the rectilinear far-field arrangement of the diffracted beam OVs that was described and substantiated in Ref. [16]. Additionally, we have demonstrated interesting topological reactions in the far-field singular skeleton evolution.

**Acknowledgements**

This work was supported, in part, by the Ministry of Education and Science of Ukraine, project No 531/15.

Video 1. Illustration of the graphical solution of Eqs. (14) - (16) (see Fig. 3a).

Video 2. Topological reaction and jump of the OV trajectory in the diffracted Kummer beam (see Fig. 3b).

Video 3. Topological reaction and jump of the OV trajectory in the diffracted LG beam: intensity pattern (see Fig. A1a).

Video 4. Topological reaction and jump of the OV trajectory in the diffracted LG beam: equiphase contours (see Fig. A1b).

Video 5. Topological reaction in the far field of the diffracted LG beam: equiphase contours (see Fig. B1b--d).